\documentclass[fleqn,usenatbib,useAMS]{mnras}

\usepackage{graphicx}	% Including figure files
\usepackage{amsmath}	% Advanced maths commands
\usepackage{amssymb}	% Extra maths symbols
\usepackage{multicol}        % Multi-column entries in tables
\usepackage{bm}		% Bold maths symbols, including upright Greek
\usepackage{pdflscape}	% Landscape pages
\usepackage{xcolor}	% more colors

\usepackage[T1]{fontenc}
\usepackage{ae,aecompl}

\usepackage{newtxtext,newtxmath}

\title[Dipolar Chromospheres in Two White Dwarfs]{Discovery of Dipolar Chromospheres in Two White Dwarfs\thanks{Dedicated to Prof.~Tom Marsh (1961--2022)}}

\author[J. Farihi et al.]{J. Farihi$^1$\thanks{E-mail: j.farihi@ucl.ac.uk},
J. J. Hermes$^2$,
S. P. Littlefair$^3$,
I. D. Howarth$^1$,
N. Walters$^1$,
S. G. Parsons$^3$
\\
$^1$Department of Physics and Astronomy, University College London, London WC1E 6BT, UK\\
$^2$Department of Astronomy \& Institute for Astrophysical Research, Boston University, Boston 02215, USA\\
$^3$Department of Physics and Astronomy, University of Sheffield, Sheffield S3 7RH, UK\\
}

\begin{document}

%\label{firstpage}

\maketitle

\begin{abstract}
This paper reports the ULTRACAM discovery of dipolar surface spots in two cool magnetic white dwarfs with Balmer emission lines, while a third system exhibits a single spot, similar to the prototype GD\,356.  The light curves are modeled with simple, circular, isothermal dark spots, yielding relatively large regions with minimum angular radii of $20\degr$.  For those stars with two light curve minima, the dual spots are likely observed at high inclination (or colatitude), however, identical and antipodal spots cannot simultaneously reproduce both the distinct minima depths and the phases of the light curve maxima.  The amplitudes of the multi-band photometric variability reported here are all several times larger than that observed in the prototype GD\,356; nevertheless, all DAHe stars with available data appear to have light curve amplitudes that increase toward the blue in correlated ratios.  This behavior is consistent with cool spots that produce higher contrasts at shorter wavelengths, with remarkably similar spectral properties given the diversity of magnetic field strengths and rotation rates.  These findings support the interpretation that some magnetic white dwarfs generate intrinsic chromospheres as they cool, and that no external source is responsible for the observed temperature inversion.  Spectroscopic time-series data for DAHe stars is paramount for further characterization, where it is important to obtain well-sampled data, and consider wavelength shifts, equivalent widths, and spectropolarimetry.

\end{abstract}


\section{Introduction}

The origin of magnetism in white dwarf stars is an outstanding astrophysical puzzle more than a half century old, but recent and ongoing developments are now shedding light on this fundamental, and still poorly understood aspect of stellar evolution.  The first signatures of white dwarf magnetism resulted from the detection of circular polarization in spectra that were quasi-featureless or with unidentified absorption bands \citep{kemp1970, angel1971, landstreet1971} that were later understood to be shifted hydrogen and (neutral) helium lines, mostly consistent with centered or offset dipole field geometries \citep{kemic1974, garstang1977, wickramasinghe1979}.  A summary of magnetic white dwarf research over the first several decades can be found in two published reviews \citep{wickramasinghe2000, ferrario2015}.

One of the key developments was the recognition that magnetic white dwarfs are nearly exclusively found as isolated stars, or in cataclysmic variables \citep{liebert2005}.  This empirical finding led to the hypothesis that fields are generated during common envelope evolution \citep{tout2008, nordhaus2011, belloni2020}, a process that may function effectively for stars, brown dwarfs, and giant planets that are engulfed during the post-main sequence \citep{farihi2011b, kissin2015, guidarelli2019}.  And while fast-spinning and massive magnetic white dwarfs are known, and thus consistent with a stellar merger origin \citep{ferrario1997b, garcia-berro2012, kilic2021, williams2022}, it is also clear that magnetism, high remnant mass, and rapid rotation are far from tightly correlated \citep{ferrario2005, brinkworth2013}.

It has been suspected for decades that cooler white dwarfs are more often found to be magnetic \citep{liebert1979, liebert2003a}.  However, luminosity and sensitivity biases exist, where the coolest white dwarfs essentially require metal pollution to detect Zeeman splitting \citep{kawka2014, hollands2015, bagnulo2019}.  Despite these uncertainties, the possibility that magnetic fields first emerge in cool and isolated white dwarfs is intriguing, as substantial cooling is necessary for core crystallization, which has been hypothesized to be a source of an internal dynamo powered by the liquid-solid phase separation at the core boundary \citep{isern2017}.  In this scenario, magnetic field generation is decoupled from external sources of mass and angular momentum, but nevertheless, all else being equal, more rapidly rotating remnants should have stronger fields.

In a pioneering effort to overcome the aforementioned biases, and determine the actual frequency of magnetism as a function of white dwarf characteristics, \citet{bagnulo2021} carried out a nearly complete census of ($N\approx150$) white dwarfs within 20\,pc.  This volume-limited survey used sensitive circular spectropolarimetry and resulted in the first unbiased study of white dwarf magnetism, where the principal findings can be summarized as follows.\\

\begin{enumerate}

\item{All spectral classes have similar incidences of magnetism, regardless of atmospheric composition.}

\item{The field strength distribution is uniform over four orders of magnitude from 40\,kG to 300\,MG.}

\item{Magnetism is detected more frequently in white dwarfs with higher than average mass.}

\item{White dwarfs with cooling ages younger than 0.5\,Gyr -- prior to core crystallization -- are rarely magnetic.}

\item{There is no evidence of field strength decay over time.}

\end{enumerate}

It is within this background of recent developments that emerged the relatively new and small class of DAHe white dwarfs (D: degenerate, A: Balmer lines strongest, H: magnetic line splitting, e: emission).  The prototype is GD\,356, an isolated $T_{\rm eff}\approx7500$\,K star with Balmer emission lines split in a $B\approx13$\,MG field.  There are deep, multi-wavelength, non-detections that yield stringent upper limits on an X-ray corona, ongoing accretion, and low-mass companions \citep{greenstein1985, ferrario1997a, weisskopf2007}.  This apparently single white dwarf has a 1.927\,h rotation period, based on a nearly sinusoidal light curve that is well modeled by single dark spot, whose size is consistent with that of the magnetic and heated region \citep{ferrario1997a, brinkworth2004}.  These enigmatic properties led to the hypothesis that, analogous to the Jupiter-Io system, the relatively cool white dwarf surface could be heated by Ohmic dissipation of a current loop set up by the orbital motion of a conducting planet \citep{li1998, wickramasinghe2010}; referred to as the unipolar inductor model.

GD\,356\footnote{Previously thought to have a helium-rich atmosphere \citep{bergeron2001, limoges2015}.} was the only known DAHe white dwarf for 35 years, until 2020 when second and third cases were reported \citep{reding2020, gaensicke2020}.  In addition to their shared spectral morphology and strong magnetism implied from Zeeman splitting, these three cool white dwarfs with emission lines all share commonalities with some magnetic white dwarfs: relatively rapid rotation, masses only slightly above average, and no evidence for low-mass stellar or substellar (detached) companions.  A detailed time-series study of the prototype has shown that (i) the spin period is stable over two decades, with no other independent frequency signals as would be expected from a unipolar inductor, (ii) the emission line strength oscillates in anti-phase with the broad-band stellar brightness, and (iii) so far, DAHe stars share a tightly correlated set of effective temperatures and luminosities \citep{gaensicke2020,walters2021}.  This clustering is potentially related to core crystallization and magnetic field diffusion toward the stellar surface \citep{ginzburg2022}.

This paper reports detailed light curves for three DAHe white dwarfs: the second known example, SDSS\,J125230.93$-$023417.7 (\citealt{reding2020}; hereafter SDSS\,J1252), and two recently identified members of this class, LP\,705-64 and WD\,J143019.29$-$562358.3 (\citealt{reding2023}; hereafter WD\,J1430).  Two of the three stars reveal light curves with asymmetric dimming events that are $180\degr$ out-of-phase, and thus consistent with dipolar star spots.  These data are inconsistent with a unipolar inductor model, and instead support the generation of intrinsic chromospheres in some isolated, magnetic white dwarfs.  The observations and data are discussed in Section 2, the time-series analysis is presented in Section 3, followed by a summary and discussion.

\section{Observations}

This study focuses on light curves and the resulting periodicities of three DAHe white dwarfs, using both ground- and space-based photometric monitoring as described below.

\subsection{Target properties and selection}

SDSS\,J1252 is the second discovered example of a DAHe white dwarf, reported to have emission lines split in a $B\approx5$\,MG field, and with a sinusoidal light curve dominated by a period of 317.3\,s \citep{reding2020}.  The fast rotation of this star makes it an attractive target for high-cadence photometric monitoring from the ground, with a goal to obtain a detailed light curve.  LP\,705-64 and WD\,J1430 are two newer members of the DAHe spectral class with indications from {\em TESS} data that their full spin cycles could each be readily covered in a single night of ground-based photometry \citep{reding2023}.  The initial observational goals were similar to those achieved by \citet{walters2021}, to establish robust ephemerides against which future period changes might be investigated (e.g.\ within a unipolar inductor and orbiting planet model), and to constrain the nature of the emitting and magnetic regions.

%%% TABLE OBSERVING RUNS %%%
\begin{table}
\begin{center}
\caption{Chronological summary of ULTRACAM observing runs\label{obssum}.}
\begin{tabular}{@{}ccccc@{}}

\hline

Target		&Observing		&$t_{\rm exp}$	&Coverage	&Filters\\
			&Dates			&(s)			&(min)		&\\
				
\hline

SDSS\,J1252	&2021 Apr 06		&10.05		&57			&$ugr$\\
			&2021 Apr 08		&10.35		&86			&$ugr$\\
			&2021 Aug 20		&10.05		&24			&$ugr$\\
	
LP\,705-64	&2021 Aug 17		&8.05		&73			&$ugi$\\
			&2021 Aug 19		&8.05		&77			&$ugr$\\
			&2021 Aug 20		&8.05		&146			&$ugr$\\

WD\,J1430	&2022 Apr 26		&6.06		&277			&$ugi$\\
			&2022 Jun 05		&6.33		&176			&$ugi$\\

\hline

\end{tabular}
\end{center}
\end{table}

\subsection{ULTRACAM observations}

%%% FIGURE 1252 RUNS %%%
\begin{figure*}
\includegraphics[width=\linewidth]{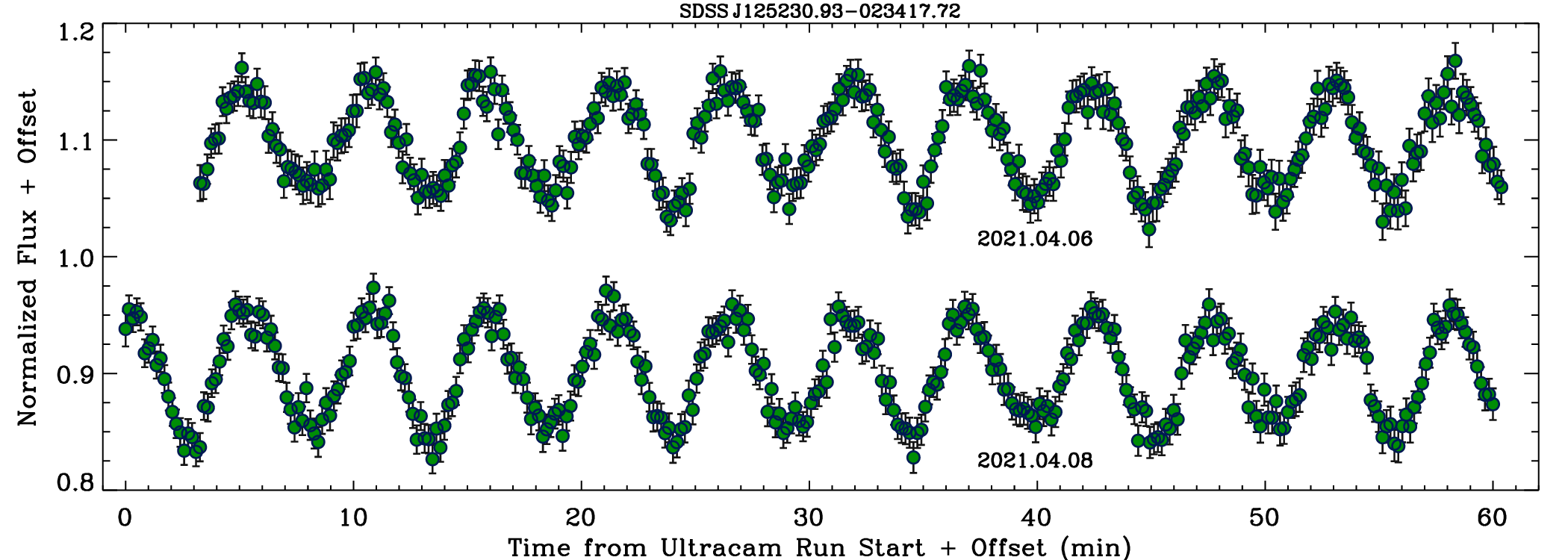}
\vskip 0pt
\caption{Approximately 1\,h of ULTRACAM $g$-band light curves for SDSS\,J1252, each taken on a different night.  The data are plotted as observed in sequence, each light curve normalized, offset vertically by $\pm0.1$, and shifted horizontally to exhibit the same photometric phase.  Visual inspection reveals that adjacent minima are unequal in depth, which is also true but more subtle for adjacent maxima.  These real-time observations were the first indication that SDSS\,J1252 has two star spots, 180\degr out-of-phase, and the true rotation period is twice as long as the 317.3\,s previously reported by \citet{reding2020}.
\label{runs1252}}
\end{figure*}

All three stars were observed with ULTRACAM, a frame-transfer CCD imaging camera (24\,ms dead time between exposures; \citealt{dhillon2007}) that is permanently mounted on the 3.6\,m NTT telescope at the La Silla Observatory in Chile.  The instrument has three independent channels that enable the use of independent filters simultaneously, and data were taken with filters similar to standard $u$, $g$, and one of $r$ or $i$ bandpasses, but with higher throughput.  In each case, the blue channel was co-added every three frames to improve the effective signal-to-noise on the target.  The observation details, including exposure times (same as the cadences for ULTRACAM), and durations of the resulting light curve segments, are summarized in Table~\ref{obssum}. 

Images were corrected for bias and flat fielded with normalized sky images obtained during evening twilight (taken in a continuous spiral to remove stars).  Differential brightnesses were measured relative to field stars with dedicated software\footnote{https://github.com/HiPERCAM/hipercam.} using photometric apertures that were typically scaled to $2\times$ the mean full width at half maximum of the stellar profiles for each exposure.  The sky annuli were fixed to span the region 8.75--15.75\,arcsec from the stars, where a clipped mean was used to determine the background.  For all stars in all observations, the same sets of comparison stars were used to generate light curves, consisting of two or three stars in the $gri$ frames, and one to two stars in the $u$-band images.

%1430 2 and 1 comps, texp 6.05695 (4 coadds) or      6.32752 (3 coadds) 
%1252 3 and 2 comps, texp 10.3542 to 10.0542, 31.1106 to  30.2106 (3 coadds) (didn't use 3rd one)
%  705 2 and 2 comps, texp 8.04778 and 3 coadds for uband (24.1914) in all observations

Light curves were constructed by dividing the science target flux by the sum of the comparison star fluxes, and normalizing the result.  Measurement errors were propagated from the aperture photometry, by summing in quadrature the fractional flux errors of all stars measured for a given light curve.  All ULTRACAM times were converted to Barycentric Julian Day (BJD) using Barycentric Dynamical Time (TDB), following \citet{eastman2010}.

\subsection{{\em TESS} data}

Data for each of the three DAHe targets are available from {\em TESS} \citep{ricker2015}, and were downloaded from the MAST archive, where the {\sc pdcsap} processed light curves were retained for analysis.  Time stamps were were corrected to BJD = {\em TESS}\_BJD + 2457000.

LP\,705-64 (= TIC\,136884288) was observed in Sector 30, while data were collected for WD\,J1430 (= TIC\,1039012860) within Sector 38, and for SDSS\,J1252 (= TIC\,953086708) during Sector 46.  All three stars have 120\,s cadence observations.  These data were further cleaned of NaN flux entries, but with no other processing based on data quality flags, yielding light curves that retained between 80 and 90 per cent of their {\sc pdcsap} array values.  Lastly, outliers beyond $\pm5\upsigma$ of the local time average (or phase average) flux were removed, which were fewer than five points in total for each source.

It is worth noting that these data are not all equally useful in subsequent analysis.  The following {\em TESS} benchmarks summarize their relative quality: SDSS\,J1252 has $G=17.5$\,mag, a mean flux of $19.2\pm5.3$\,e$^-$\,s$^{-1}$ (28 per cent scatter); LP\,705-64 has $G=16.9$\,mag, a mean flux of $38.0\pm5.6$\,e$^-$\,s$^{-1}$ (15 per cent scatter), while WD\,J1430 has $G=17.4$\,mag, a mean flux of $9.6\pm5.4$\,e$^-$\,s$^{-1}$ (73 per cent scatter), and lies within the Galactic plane.

\section{Time-series analysis and results}

All light curves were analyzed using {\sc period04} \citep{lenz2005}, where a Lomb-Scargle periodogram was constructed using ULTRACAM data, {\em TESS} {\sc pdcsap} light curves, or a combination of the two datasets, with a goal to identify that which produces the most precise ephemerides for each target.  Monte Carlo simulations, run within {\sc period04}, were used to determine errors in frequency and phase for the strongest periodogram peak for each star and set of light curves, then propagated to determine the error in $T_0$ corresponding to photometric minimum.  The frequency and phase were allowed to vary independently during the simulations to determine errors, which were typically repeated 1000 times.

\subsection{SDSS\,J1252}

For SDSS\,J1252, there are sufficient ULTRACAM data to uniquely determine the photometric period and provide an improved ephemeris.  Light curves cover more than 30 epochs of its previously reported 317.3\,s periodicity (frequency 272.3\,d$^{-1}$, \citealt{reding2020}), with an observational baseline of 136\,d, spanning several $10^4$ cycles at this frequency.  In Figure~\ref{runs1252} are shown the first and second $g$-band light curves obtained for this white dwarf, from which can be discerned that there are {\em two distinct set of minima (and maxima)}, each manifesting every 317.3\,s, and thus revealing an actual photometric period of 634.6\,s.

\begin{figure}
\includegraphics[width=\columnwidth]{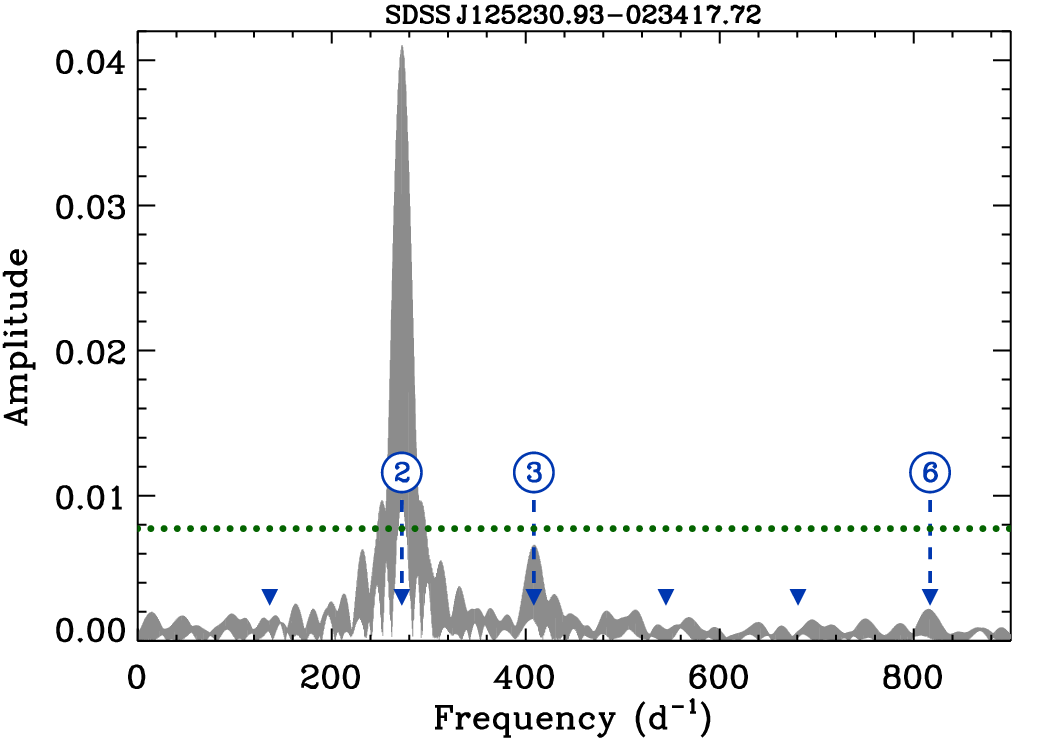}
\vskip 0pt
\caption{Periodogram of SDSS\,J1252 based on three nights of ULTRACAM data using co-added, $g+r$-band light curves, with amplitudes plotted in grey.  The data were bootstrapped 10\,000 times to determine the amplitude above which represents a 0.1 per cent chance a signal is spurious.  This false alarm amplitude of 0.007734 is delineated by the green dotted line, where only the strongest peak is higher. However, the two frequencies with the largest periodogram peaks have a near exact ratio of 3:2, and a weaker third peak is consistent with a frequency that is an integer multiple of both lower frequencies.  For a fixed rate of stellar rotation, and despite a lack of significant periodogram power (amplitude), this result indicates the fundamental frequency is 136.150\,d$^{-1}$.  This is half of the frequency with the largest periodogram signal, and is consistent with two distinct, out-of-phase star spots tracing the observed light curve morphology.  The first six harmonics of the fundamental are marked with blue triangles, where only those showing noteworthy amplitude are numbered.  
\label{pgram1252}}
\end{figure}

The ULTRACAM $g+r$ co-added light curves were analyzed using data from all three observing runs.  The resulting best periodogram is plotted in Figure~\ref{pgram1252}, where the strongest peak is identical to the frequency reported in the discovery paper; however, there is a second outstanding signal near 408\,d$^{-1}$.  This secondary peak is not as well-determined as the 272.3\,d$^{-1}$ signal, but these two frequencies appear to have a near an exact ratio of 3:2.  Additionally, the periodogram also reveals a weak-amplitude peak at roughly 816\,d$^{-1}$, and the ratio of these three frequencies is 6:3:2.

%%% FIGURE 1252 PGRAM %%%

These periodicities are far shorter than any possible range of orbital signals originating from non-degenerate companions, as the lowest-frequency periodogram signal, at 272.3\,d$^{-1}$, corresponds to a Keplerian orbit near $7\,R_\star$ (seven white dwarf radii), deep within the nominal Roche limit.  Only a compact object could survive at this orbital distance, such as those in close but detached, double white dwarf binaries.  And while there are a few such systems known to have orbital periods comparable to the frequencies exhibited by SDSS\,J1252, their light curves reveal ellipsoidal modulation owing to tidal distortions in the primary white dwarfs \citep{kilic2011, brown2011b, burdge2019}.  Furthermore, these rare, deformed degenerates are all helium-core white dwarfs less massive than 0.3\,M$_{\odot}$, and thus significantly more prone to tidal distortion than SDSS\,J1252 and the DAHe stars, which are considerably more compact \citep{walters2021, reding2020, gaensicke2020}.

Therefore, the light curve and resulting periodogram of SDSS\,J1252 are interpreted as arising from a single star.  It is reasonable to assume the $T_{\rm eff}\approx8000$\,K white dwarf has a fixed spin period (no differential rotation), and no stellar pulsations, as it is far from the hydrogen atmosphere instability strip \citep{romero2022}.  The observed signals are then interpreted as the 2$^{\rm nd}$, 3$^{\rm rd}$, and 6$^{\rm th}$ harmonics of the stellar rotation frequency.  The periodogram signals and their amplitudes reflect the fitting of sinusoids to the light curve, where the two highest have a 3:2 ratio in order to generate both the principal flux variation at 272.3\,d$^{-1}$, and the alternating minima via the interference with the 408\,d$^{-1}$ frequency.  

The revised stellar rotation period and associated uncertainty were determined by dividing the 2$^{\rm nd}$ harmonic frequency by two, yielding 136.15032(2)\,d$^{-1}$ (equivalent to a period of 634.59273(9)\,s).  Although essentially no amplitude (or power) is seen in the periodogram at the frequency inferred to be the fundamental, this is an expected consequence of the light curve morphology and sinusoidal fitting \citep{vanderplas2018}.

A similar analysis was attempted using the {\em TESS} light curve, both on its own and in combination with ULTRACAM data.  While the {\em TESS} 120\,s cadence is $2.6\times$ faster than the peak periodogram frequency for SDSS\,J1252, and thus above the Nyquist rate, the data quality are relatively poor (Section~2.3).  No time-series analysis utilizing {\em TESS} led to any improvement in frequency or phase precision, and therefore all calculations for SDSS\,J1252 are based solely on the ULTRACAM observations.

\subsection{LP\,705-64 and WD\,J1430}

\begin{figure*}
\includegraphics[width=\linewidth]{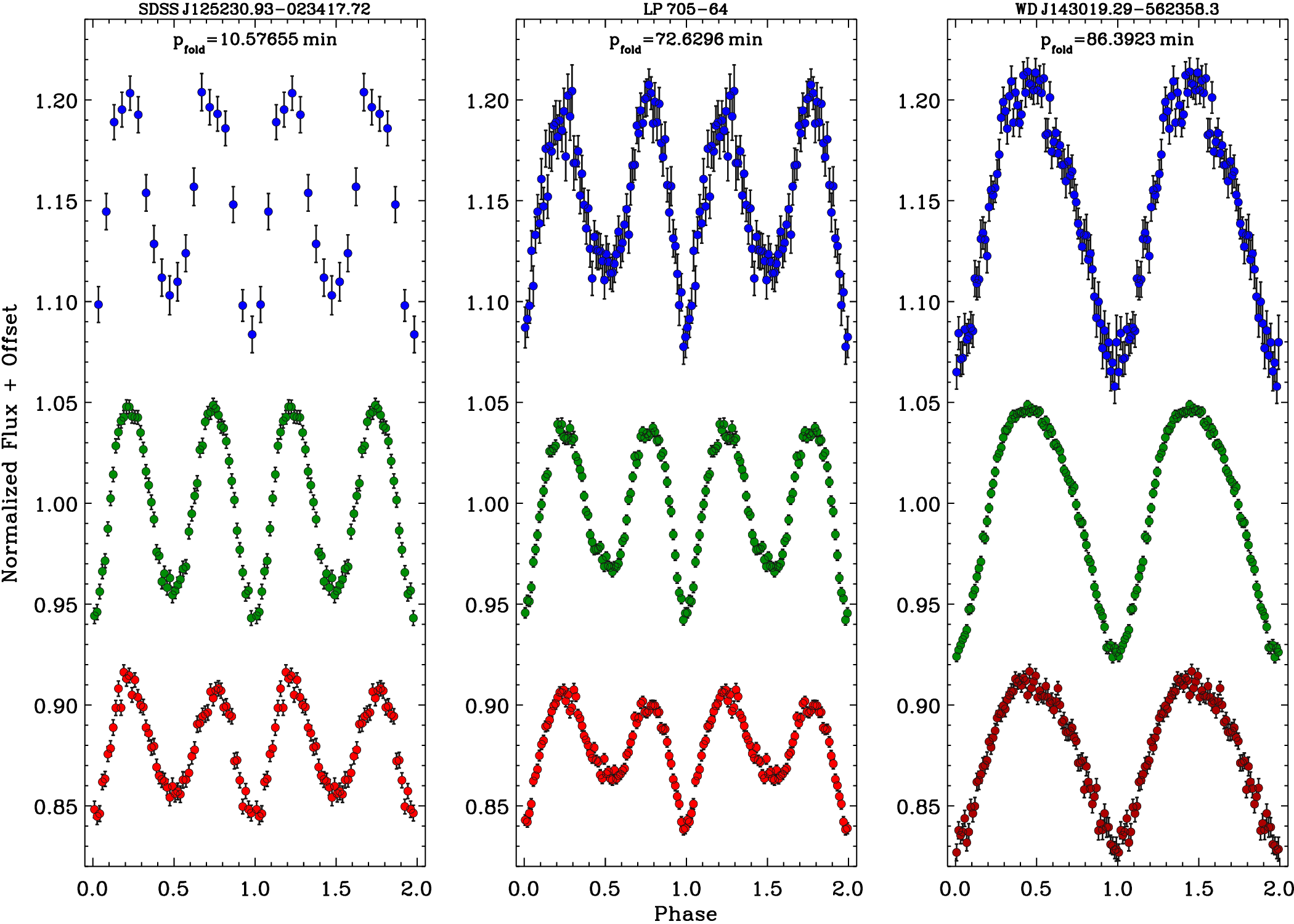}
\vskip 0pt
\caption{Normalized and phase-folded ULTRACAM light curves for all three stars, where the blue points are $u$ band, the green points are $g$ band, and the red points are $r$ or $i$ band.  The time-series data have been folded on the periods listed at the top of each panel, and have been re-sampled onto regular grids.  There are 80 phase bins for LP\,705-64 and WD\,J1430 in all three filters, while for SDSS\,J1252 there are 60 phase bins for $g$ and $r$, but only 20 for the $u$ band (Section~3.3).  These light curves highlight the asymmetric, anti-phase modulation from two starspots in the case of both SDSS\,J1252 and LP\,705-64. 
\label{lcfigs}}
\end{figure*}

{\em TESS} data were the initial means of identifying the stellar rotation rates in these two DAHe white dwarfs \citep{reding2023}.  However, similar to as observed in SDSS\,J1252, the ULTRACAM light curve of LP\,705-64 exhibits two unequal minima in a single cycle, and thus the period determined by {\em TESS} represents one half its spin period (see Figure~\ref{lcfigs}).  For this source, the ULTRACAM data alone do not span a sufficient number of cycles to determine the photometric period with precision comparable to {\em TESS}.  A significant improvement in the {\em TESS} ephemeris is achieved using the combination of ULTRACAM $g+r$ co-added light curves and {\em TESS}, resulting in a periodogram with a single peak at 39.65325(3)\,d$^{-1}$ [cf.\ 39.653(2)\,d$^{-1}$; \citealt{reding2023}], and a corresponding higher precision in phase.  However, the true spin period must be calculated from this frequency by recognizing it is the 2$^{\rm nd}$ harmonic of the fundamental, which is 19.82662(1)\,d$^{-1}$.

For WD\,J1430, the ULTRACAM light curves reveal a single maximum and minimum with one of the largest amplitudes observed to date for a DAHe white dwarf (5.8 per cent in the $g$ band).  Similar to LP\,705-64, there are insufficient ULTRACAM data from which to derive a precise ephemeris for this source, and thus the combination of ULTRACAM and {\em TESS} Sector 38 {\sc pdcsap} data were utilized for this goal.  Initially, the analysis of these combined datasets improved the precision of the periodogram frequency, but resulted in phase errors that were larger than those based on {\em TESS} alone.  Subsequently, these {\em TESS} data were re-scaled (see Section~3.4) to more closely match those of the co-added $g+i$-band ULTRACAM data, and the resulting analysis marginally improved the uncertainty in phase.  Ultimately for this star, the best constraints were achieved by adding a third set of light curves into the periodogram analysis, using full-frame data from {\em TESS} Sector 11, where fluxes were extracted based on PSF-subtracted images following \citet{han2023}.

%%% FIGURE ALL LCS %%%

\subsection{Light curve morphologies, ephemerides, spectral phases}

Based on the preceding analysis, and to reveal light curve structures more precisely as a function of phase, the ULTRACAM multi-band data were phase-folded and binned using a weighted average onto regular grids.  The resulting light curves are shown in Figure~\ref{lcfigs}, where there are 80 phase bins for LP\,705-64 and WD\,J1430 in all three channels.  In the case of SDSS\,J1252, the short spin period indicates that a single ULTRACAM frame has a phase width of 0.016 in the green and red channels, but 0.048 in the blue channel (owing to three co-adds).  For this reason, the light curves of SDSS\,J1252 were re-sampled into 60 phase bins in $g$ and $r$, but only 20 bins in $u$ band.

The folded light curves for SDSS\,J1252 and LP\,705-64 both exhibit alternating minima that are indicative of two distinct star spots $180\degr$ out-of-phase during rotation.  While this behavior is not novel among magnetic white dwarfs, there appear to be only a few documented examples of white dwarf light curves where dipolar spots are suggested or required \citep{hermes2017, kilic2019, pshirkov2020}.  In contrast, the majority of magnetic white dwarf light curves seem to be broadly consistent with sinusoidal (single spot) morphologies \citep{brinkworth2013}, including the prototype DAHe star GD\,356 \citep{walters2021}.  However, it should be noted that incomplete phase coverage and modest photometric precision can inhibit the detection of subtle light curve features (e.g.\ the discovery light curve of SDSS\,J1252, and the {\em TESS} light curve of LP\,705-64; \citealt{reding2020, reding2023}). 
		
To calculate accurate ephemerides based on the best precision achieved here, $T_0$ was chosen from an ULTRACAM light curve located nearest to the middle of the temporal coverage for each star, and where a feature could be unambiguously identified as a true photometric minimum.  The periodogram analysis of the preceding sections then results in the following best ephemerides for all three DAHe white dwarfs, where zero phase corresponds to actual photometric minimum, and the periods are accurate and precise determinations of their spins:

\begin{flushleft}
\setlength\tabcolsep{1pt}
\begin{tabular}{@{}lll@{}}

BJD$_{\rm TDB}$ (SDSS\,J1252) 	&= 2459313.809921(6)     &+ 0.007344823(1)\,$E$\\
BJD$_{\rm TDB}$ (LP\,705-64) 	&= 2459444.92339(6)   	&+ 0.05043723(3)\,$E$\\
BJD$_{\rm TDB}$ (WD\,J1430)		&= 2459696.8239(3)     	&+  0.05999529(3)\,$E$\\

%BJD$_{\rm TDB}$ (WD\,J1430)	&= 2459696.8240(6) 	&+  0.0599952(1)\,$E$\\

\end{tabular}
\end{flushleft}

\noindent
As mentioned earlier, the ephemeris of SDSS\,J1252 is based solely on ULTRACAM, whereas those of LP\,705-64 and WD\,J1430 are based on the combination of {\em TESS} and ULTRACAM.  From these ephemerides, forward and backward extrapolations can be made and compared with published, time-varying spectra of LP\,705-64 and WD\,J1430, but in the case of SDSS\,J1252 there is insufficient time resolution to compare its spectroscopic variations in phase with photometry \citep{reding2020,reding2023}.

Starting with the more straightforward case of WD\,J1430 which exhibits a single spot, the photometric minimum (phase 0) occurred at BJD$_{\rm TDB}=2459049.056\pm0.001$ nearest the two epochs of the published spectroscopy.  This notably falls close to halfway in time between the two spectra plotted and described by \citet{reding2023} as 'emission' and 'absorption'.  Specifically, and taking the reported epochs of observation at face value, these spectra correspond to photometric phases $0.720\pm0.007$ and $0.221\pm0.007$, respectively, and thus both occur close to the average stellar flux.  While these two spectral phases are reported as potentially representing a maximum and minimum magnetic field strength, this interpretation seems uncertain, especially if other spectral phases exhibit weaker emission or absorption, where Zeeman splitting is not a robust diagnostic.

Superficially interpreting these spectral phases of WD\,J1430 as the highest and lowest field strength would be somewhat inverse to that observed for the prototype DAHe GD\,356, where there are multiple, full-phase coverage observations using both spectroscopy and spectropolarimetry.  For this well-studied case, the magnetic field variations, both from the observed parallel component and using Zeeman splitting, display a peak and trough near phases 0.3 and 0.8, respectively, from photometric {\em minimum} \citep{walters2021}.  For WD\,J1430, existing data may not probe the magnetic field with sufficient sensitivity or phase coverage, and hence these comparative results should be considered preliminary at best.

In the case of LP\,705-64, the situation is more complex.  Depending on the spot sizes, one might expect {\em two minima and maxima in both equivalent width and magnetic field variations, one pair associated with each spot}.  However, there are only two epochs of spectroscopy plotted and described by \citet{reding2023}, and here again a comparison must be considered not only preliminary, but possibly inapt for the aforementioned reasons.  Again taking the published epochs at face value, and where the the deeper of the two light curve minima is zero phase, the spectrum shown with the broader Zeeman splitting corresponds to photometric phase $0.048\pm0.001$.  The two reported spectral epochs were chosen as to be separated by exactly one half spin cycle \citep{reding2023}, so that further interpretation would reflect the selection.

While the updated photometric ephemeris is sufficient to predict precise spin phases for spectroscopic observations of LP\,705-64, their potential correlation is not yet straightforward.  It has not yet been demonstrated that high and low Zeeman splitting might be in-phase with photometric extrema (cf.\ GD\,356 \citealt{walters2021}).  The two published spectra may not represent precise peak behavior, and there may be some uncertainty in the epoch dates reported.  Measurements of both equivalent width and magnetic field strength at all rotational phases would eliminate these ambiguities.  The sparse set of published spectroscopic measurements of DAHe white dwarfs, currently prevents a more robust correlation of photometric and spectroscopic phases.

%%% TABLE RESULTS %%%
\begin{table}
\begin{center}
\caption{Multi-wavelength variability amplitudes $A_\uplambda$ in per cent flux\label{lcamp}.}
\begin{tabular}{@{}rrrr@{}}

\hline

SDSS\,J1252		&LP\,705-64		&WD\,J1430				&GD\,356\\

\hline

$u$: $5.72\pm0.23$	&$u$: $4.48\pm0.17$	&$u$: $6.62\pm0.14$	&$u$: $1.50\pm0.06$\\
$g$: $4.97\pm0.06$	&$g$: $3.92\pm0.06$	&$g$: $5.78\pm0.04$	&$g$: $1.22\pm0.04$\\
$r$: $2.99\pm0.06$	&$r$: $2.38\pm0.05$		&...					&$V$+$R$: $0.81\pm0.02$\\
...				&...					&$i$: $3.84\pm0.07$		&...\\
$T$: $2.05\pm0.30$	&$T$: $1.60\pm0.17$	&...					&$T$: $0.62\pm0.02$\\

% used gr.dat for wd1832 to get f= 489.888813 and use this to get ugr amplitudes
% u 	4.87 +- 0.26  -- > u/g = 1.15
% g 	4.22 +- 0.13
% r  	3.85 +- 0.18. --> r/g = 0.91 **

% used g.dat for wd1859 to get f= 207.232726 and use this to get ugr amplitudes
% u 	1.24 +- 0.06  -- > u/g = 1.05 **
% g 	1.18 +- 0.02
% r  	0.56 +- 0.04. --> r/g = 0.47 **

\hline

\end{tabular}
\end{center}
{\em Note}.  The $u$-band amplitude for GD\,356 was obtained from a light curve taken at the WHT using PF-QHY on 2020 Jun 21.  $A_T$ is the flux variation amplitude in the {\em TESS} band, where is no corresponding entry for WD\,J1430.  While both SDSS\,J1252 and WD\,J1430 are similarly faint ($G\approx17.5$\,mag) and thus near the limit of what {\em TESS} can observe, the latter source has a target-to-total aperture flux of only 0.03 (cf.\ 0.89 for SDSS\,J1252, 0.56 for LP\,705-64, and 0.87 for GD\,356; \citealt{walters2021}).  Thus the {\em TESS} amplitude for WD\,J1430 is likely unreliable.
\end{table}

% WDJ1430
% CROWDSAP=         0.03215283 / Ratio of target flux to total flux in op. ap.  
% FLFRCSAP=           0.57686555 / Frac. of target flux w/in the op. aperture  

%		THIS MAY MEAN THE EPHEMERIS IS BULLSHIT TOO

% LP705-64
% CROWDSAP=         0.55918050 / Ratio of target flux to total flux in op. ap.  
% FLFRCSAP=           0.65971649 / Frac. of target flux w/in the op. aperture  

% SDSS1252
%CROWDSAP=           0.89320248 / Ratio of target flux to total flux in op. ap.  
%FLFRCSAP=           0.38091946 / Frac. of target flux w/in the op. aperture  

% GD356
% an average ratio of target flux to total flux (CROWDSAP) in the target aperture 0.87 over five sectors.

\subsection{Multi-band light curve amplitudes}

%%% FIGURE PHOT AMP RATIOS %%%
\begin{figure}
\includegraphics[width=\columnwidth]{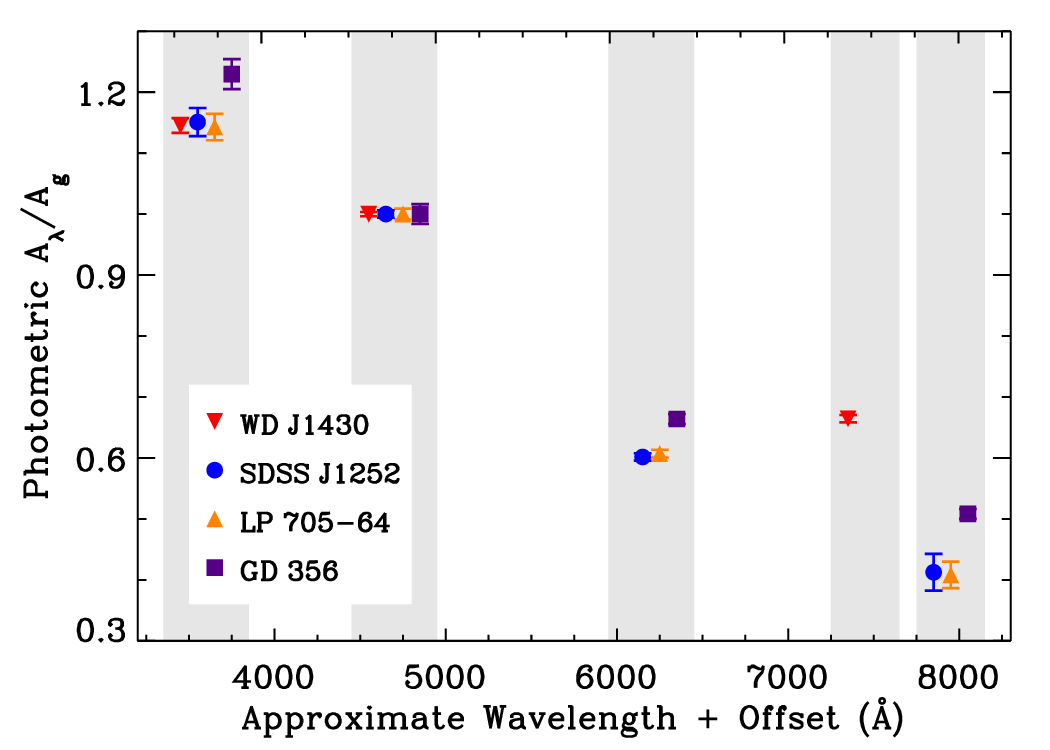}
\vskip 0pt
\caption{The multi-band photometric variability amplitudes of the three DAHe white dwarfs with ULTRACAM light curves, together with similar measurements for the prototype GD\,356 \citep{walters2021}.  The adopted central wavelengths are 3600, 4700, 6200, and 7500\,\AA \ for $ugri$ \citep{fukugita1996,gunn2006}, and 7900\,\AA \ for {\em TESS} \citep{ricker2015}. All data were analyzed in a uniform manner for this plot and the Table~\ref{lcamp} values, using {\sc period04} as described in Section~3.4.  The light curve amplitudes are normalized to the $g$-band value for a given star, with a horizontal offset of $\pm100$\,\AA \ applied to separate the data points, and errors given for the sinusoidal fits to individual bandpass data.
\label{varamp}}
\end{figure}

To better understand the nature of the spots and their associated magnetic regions, multi-band light curves for DAHe white dwarfs were used to calculate the photometric amplitude for each star in each observed bandpass.  For the three stars observed by ULTRACAM as well as GD\,356, this was done by taking only the {\em strongest} signal in the best periodogram for each star (Figures~2 and 3), and determining the sinusoidal amplitude for each light curve in each bandpass at that frequency.   In this way, all light curve amplitudes are evaluated by their strongest sinusoidal components, including those stars with little or no periodogram power at their true rotation frequency.  Light curve amplitudes and uncertainties were determined using {\sc period04}, using fixed frequencies and Monte Carlo simulations.

Table~\ref{lcamp} lists the multi-band, photometric amplitudes for the four DAHe white dwarfs with available data, the three stars reported here and the prototype GD\,356 \citep{walters2021}.  It is interesting to note that the amplitude of photometric variation is relatively small in GD\,356 (on the order of 1 per cent; \citealt{brinkworth2004}), compared to the newer DAHe stars with several percent variations in their light curves.  Although there is not yet any published multi-band photometry of SDSS\,J1219, its light curve amplitude is around 3 per cent in the $B$ band (\citealt{gaensicke2020}; roughly halfway between the $u$ and $g$ filters), and thus at least double that of GD\,356 at similar wavelengths.  It is also possible that WD\,J161634.36+541011.51 (\citealt{manser2023}; hereafter WD\,J1616) has a photometric amplitude comparable to the strongest found here.  These are likely the results of observational bias, which enhances their detection as variables in surveys such as {\em Gaia} and ZTF (e.g.\ \citealt{guidry2021}).

It should be noted that WD\,J1430 is positioned in a crowded field at Galactic latitude $\upbeta<4$\,degr, and the {\em TESS} fluxes are dominated by other sources in the photometric aperture (pipeline keyword {\sf crowdsap} = 0.032).  This pipeline metric implies that only 3.2 per cent of the flux in the extracted aperture is likely from the white dwarf, and subsequently the extracted flux has been dramatically reduced to recover a more accurate stellar brightness in the {\sc pdcsap} light curve \citep{stumpe2012}.  While the ULTRACAM observations independently confirm the stellar spin frequency identified by periodogram analysis of the {\em TESS} light curve, the pipeline fluxes are simply too noisy (see Section~2.3) and likely offset significantly from the true mean flux.  Thus, no reliable variability amplitude can be deduced for the {\em TESS} bandpass (see footnote to Table~\ref{lcamp}).  

The relative strengths of the photometric variations in DAHe stars appear to follow a trend as a function of wavelength, with increasing amplitudes towards the blue.  Figure~\ref{varamp} plots the strengths of the multi-band variability for the four white dwarfs, where the photometric amplitude for each bandpass is plotted relative to the $g$ band for each star.  Three of the four stars have data in $ugr$ (or similar) bandpasses, and all three exhibit a relatively tight correlation in their amplitude ratios as a function of these three wavelength ranges.  Three stars have reliable {\em TESS} amplitudes where again the same behavior is evident, and suggesting a phenomenon associated with this emerging spectral family.   Based on the narrow range of $T_{\rm eff}$ among DAHe white dwarfs, Figure~\ref{varamp} implies their spots have similar spectral properties.  This indication is remarkable given the range of DAHe rotation periods and especially magnetic field strengths.

%\jf{Is the pattern is associated with all magnetics in this part of the HRD, or only DAHe?  Any ULTRACAM fore DAH-no-e stars in this region?  Steven and Stu?}

To date, only relatively weak Balmer features have been detected blueward of H$\upbeta$ in any DAHe star, and yet the photometric variability remains strongest at shorter wavelengths.  This is consistent with the previous finding that the photometric variability arises from changes in the stellar continuum, and not from fluctuations in the Balmer emission lines \citep{walters2021}.

%\jf{Spot sizes, or more likely contrast, may change but the emission lines don't?}
In $2002-2003$, the $V$-band (5500\,\AA) light curve amplitude of the GD\,356 was recorded as 0.2 per cent \citep{brinkworth2004}.  But in 2020 the photometric variations observed using an SDSS $g$-band filter (4700\,\AA), and a $V+R$ filter (6200\,\AA) were found to be $4\times$ $-$ $6\times$ higher (Table~\ref{lcamp}).  It thus seems possible that the starspot has evolved during this time frame; however, all six emission features within H$\upalpha$ and H$\upbeta$ seem consistent over at least 35 years \citep{greenstein1985, ferrario1997a, walters2021}.

\subsection{Simple spot modeling}

\begin{table}
\begin{center}
\caption{Spot modeling parameter ranges and step sizes\label{spotp}.}
\begin{tabular}{@{}ccc@{}}

\hline

Parameter		&Range			&Step size\\

\hline

$i$			&$10\degr-90\degr$	&$10\degr$\\
$\uptheta$	&$10\degr-90\degr$	&$2\degr$\\

$\upalpha_1$	&$20\degr-80\degr$ 	&$5\degr$\\
$\upalpha_2$	&$5\degr-20\degr$ 	&$5\degr$\\

$f_1(T)$		&0.99-0.48 		&0.01\\
$f_2(T)$		&0.99-0.10 		&0.01\\

\hline

\end{tabular}
\end{center}
\end{table}

%%% FIGURE SPOT MODELS %%%
\begin{figure}
\includegraphics[width=\columnwidth]{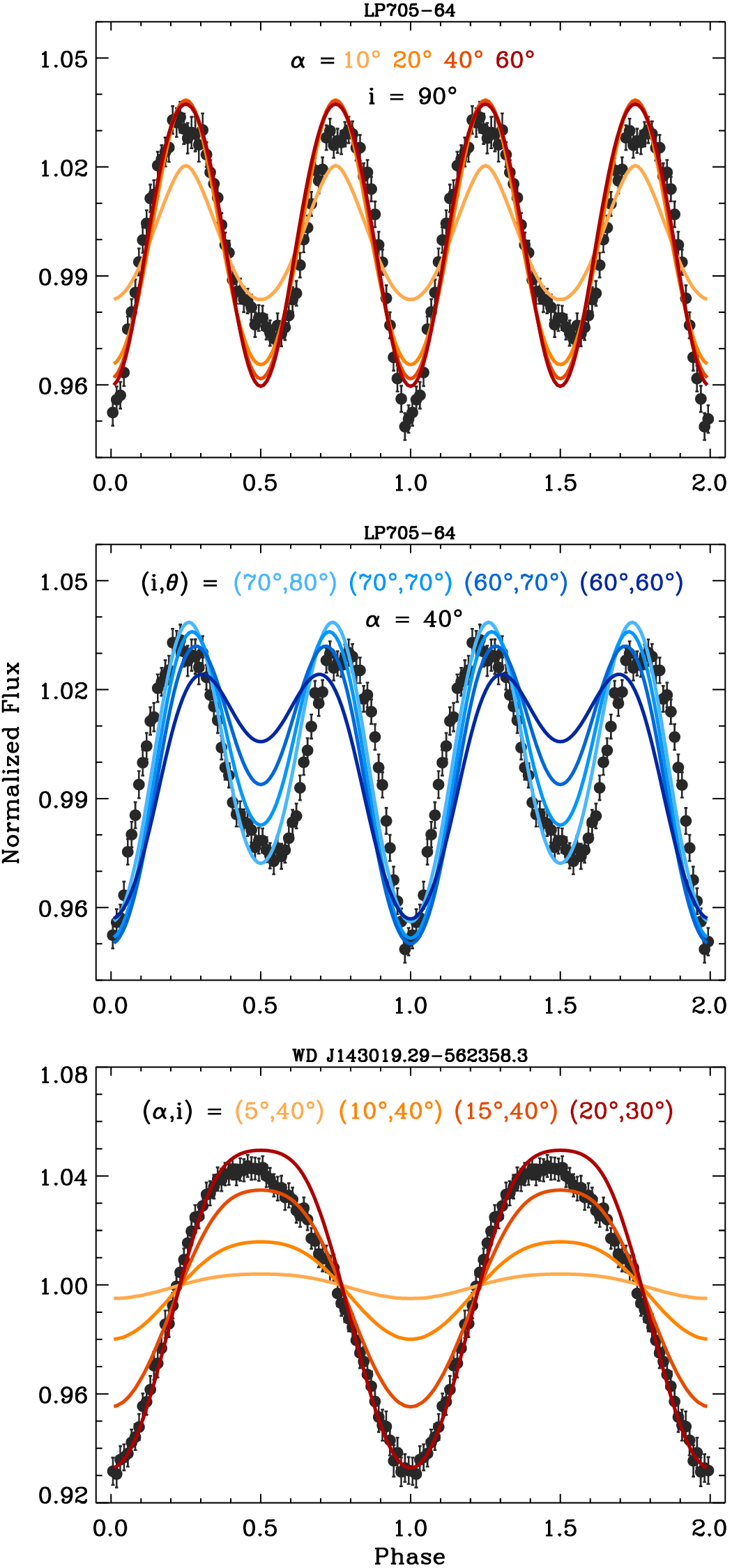}
\vskip 0pt
\caption{Illustrative spot models fitted to ULTRACAM light curves.  The upper panels plot the $g+r$-band data for LP\,705-64, which exhibits the more extreme depth change between its two minima (cf.\ SDSS\,J1252 in Figure~\ref{lcfigs}), and for which the simple models calculated here are moderately deficient.  In the top panel are shown the RMS difference minimum models over a broad range of (fixed) spot sizes, all of which result in the highest inclination (or colatitude).  The reason these maximum inclination models fit the data best is illustrated, by contrast, in the middle panel, where a representative model at lower inclinations is shown for a (fixed) spot radius of $40\degr$.  In these middle panel models, the resulting secondary minima are decreasing in depth, but this causes an increasing shift in the phase positions of the predicted flux maxima (toward phase 0.5).  The bottom panel plots the $g+i$-band light curve of WD\,J1430 with analogous models that demonstrate a spot radius smaller than around $20\degr$ is insufficient to reproduce the observed variations.
\label{spots}}
\end{figure}

A basic set of spot models and corresponding light curves were constructed to better constrain the observed stellar surfaces as a function of rotation, with a particular motivation towards those stars where two spots appear to be necessary.  Each white dwarf was treated as a $T_{\rm eff}=8000$\,K blackbody, with one or two circular, isothermal spots whose single temperature is controlled by a scaling factor $f(T)$.  Where required by the light curve morphology, two identical spots were placed on the surface at antipodal points.  The other model parameters are the inclination ($i$) of the stellar rotation axis to the observer, the spot colatitude ($\uptheta$), and the spot angular radius ($\upalpha$).  It is commonly acknowledged that such models are potentially degenerate when the values of $i$ and $\uptheta$ are interchanged (e.g.\ \citealt{wynn1992}).

For each star, a grid of models was generated with the following parameter ranges and step sizes as given in Table~\ref{spotp}.  For small angular radii ($\upalpha<20\degr$) the spot temperature range was expanded because in order to reproduce a fixed photometric amplitude, the smallest spots must be darkest.  The root mean square (RMS) difference between the model and observed fluxes in a given bandpass as a function of spin phase was computed and used to identify a best-fit model for each $\upalpha$, although in practice there is nearly always a range of models that yield similarly satisfactory results.

While the modeling is relatively simple, a few basic results emerge.  Small spots with $\upalpha\lesssim15\degr$ cannot generate a sufficiently large photometric amplitude for any of the three white dwarfs with ULTRACAM data, but otherwise the spot size is mostly unconstrained.  Beyond this small angular size threshold, the combination of adjustable geometry and spot temperature permits sufficient model flexibility to achieve comparably good fits within a range of the other three parameters.  Nevertheless, the spots must be at least modestly large; in terms of solid angle, covering several per cent or more of one hemisphere.  In the case of WD\,J1430, which has the largest photometric variability amplitude, only spots with $\upalpha\geq20\degr$ can reproduce the observed flux changes.

However, the models with the smallest RMS differences for the light curves with two minima exhibit clear shortcomings.  The resulting inclinations (or colatitudes) tend toward $90\degr$, and consequently the model fits have equally deep, light curve minima.  These fitted parameters are driven in the direction of maximum inclination (or colatitude) because, while the lower $i$ (or $\uptheta$) solutions can reproduce a more shallow secondary minimum as observed, this particular shape exhibits light curve maxima whose phase positions are shifted towards the secondary minimum.  Examples of these modeling outcomes are illustrated in Figure~\ref{spots}.

Despite these limitations, the modeling demonstrates that the basic geometry of antipodal spots is essentially correct.  However, in the context of the simple model assumptions, it is unlikely that there is a centered but tilted, symmetric dipolar arrangement for the dual-spotted stars.  Instead, the spots may not be circular, and where two opposing spots are necessary each may be distinct in size, shape, or temperature; alternatively, the spots may be in a dipolar configuration that is offset from the rotational center of the star.

%\jf{wondering if smaller spots will work for gd356 because the variability amplitude is only 1 per cent?  that would mean it would work for other DAHe... then the lesson here is either 1) large spots are always present, but rarely coming in/out of view, or 2) a variety of spot sizes exist?}.

\section{Discussion and Summary}

The discovery of DAHe white dwarfs whose light curves require two spots in a basic dipolar configuration, now totaling at least three systems \citep{manser2023}, is a modest breakthrough in their characterization.  It raises the immediate question of whether all DAHe stars have dual spots, which manifest as light curves with either one or two minima, depending on the viewing angle and spot orientation.  As of this publication, there are now just over two dozen known DAHe stars, but where only six have robustly measured light curves \citep{reding2020, gaensicke2020, reding2023, manser2023}.  There is a seventh DAHe candidate with a well-measured {\em TESS} light curve, SDSS\,J041246.85+754942.26, but which currently lacks any type of magnetic field indication or upper limit \citep{tremblay2020, walters2021}.  

Of these seven objects, three of their light curves exhibit two photometric minima, and four are consistent with a single minimum.  With such small numbers and weak constraints on spot properties, a statistical assessment of the inferred viewing geometry is not possible, but the data to date are likely consistent with all class members having dipolar magnetic and spotted regions.  If the simple modeling performed here is any indication, it may be that magnetic (spot) axes must be highly inclined towards the viewer (or equivalently have similar colatitudes) for both spots to transit, i.e.\ have ingress and egress as opposed to being partly visible at all times.

The detection of photometric variability in GD\,356 yielded limited results on its spot properties, where the size of the temperature-contrast surface was assumed to be identical to that of the magnetic region inferred from modeling of spectropolarimetry, around $40\degr$ \citep{brinkworth2004}. Otherwise the modeling followed the same assumptions as those described in Section~3.5, and the mostly sinusoidal light curve was ultimately fitted to two sets of models, one with a dark spot near the rotational pole (low $\uptheta$) viewed at high inclination, and a second viewed near the axis of rotation (low $i$), but with high colatitude; an example of the degeneracy between $i$ and $\uptheta$.  In the case that GD\,356 has antipodal spots, in the former scenario the secondary spot can remain hidden from the observer at all spin phases, and in the latter scenario, it is possible for both spots to be partly visible at all times.  If the prototype does indeed have two spots, the previous photometric modeling would disfavor the latter orientation, as it would result in some light curve impact on from both spots.

As with the DAHe prototype, it is tempting to co-identify the sizable spots with their magnetic and chromospherically active regions \citep{ferrario1997a, brinkworth2004}.  In one such picture, the spots are dark and magnetic regions underlying the chromospheric activity, so that the Balmer emission lines are at maximum brightness when the stellar continuum yields photometric minimum \citep{walters2021}.  This behavior may also be seen in SDSS\,J1212 and SDSS\,1219 \citep{reding2020, gaensicke2020}, but insufficient phase coverage and sampling prevent any certainty at present, and equivalent widths have not been determined for those stars.  The results here for LP\,705-64 and WD\,J1430 are currently ambiguous for similar reasons, and owing to additional complications.

Interestingly, time-series spectroscopy for SDSS\,J1252, WD\,J1430, and WD\,J1616 suggest that their emission lines may effectively disappear at some phases \citep{reding2020, reding2023, manser2023}, presumably when a spot or spots (and associated magnetic region) are out of view or have minimum visibility.  However, as discussed in Section~3.3, this is likely an oversimplified picture; with the exception of GD\,356, there is a distinct lack of magnetic field determinations across the entire spin phases of DAHe white dwarfs, and Zeeman splitting may be an ineffective tool for weak or transient emission features.

At present, empirical metrics associated with published, DAHe time-series spectroscopy are sparse, and it would be ideal for observers to provide both equivalent widths and central wavelengths for emission features over at least one full cycle with sufficient sampling.  In contrast to magnetic field strength estimates, which may not be possible to measure at all spin phases via Zeeman splitting if magnetic regions rotate in and out of view, only the phase behavior of equivalent width has been robustly characterized, and only in the prototype \citep{walters2021}.  It is thus essential that full spectroscopic phase coverage of DAHe white dwarfs is carried out with these measurements in mind, and where spectropolarimetry will be more sensitive to magnetic field strength, particularly when emission or absorption features are weak.

The dual-spotted nature of at least three DAHe white dwarfs has direct bearing on the hypothesis that a heated region can be caused by star-planet interactions such that a current loop is dissipated in one region of the star (e.g.\ the unipolar inductor \citealt{li1998, wickramasinghe2010}).  If such planetary interactions are in fact taking place within the strong magnetospheres of DAHe white dwarfs, they are unlike the interactions that lead to unipolar, Jupiter-Io footprint mechanisms \citep{goldreich1969}.  Going forward, models that require the presence of closely orbiting and interacting planets, which at present lack no empirical support in observations of DAHe stars, should require strong evidence to be re-considered.  Given the lack of additional periodic signals and the compelling evidence of DAHe white dwarf clustering in the HR diagram \citep{walters2021, reding2023, manser2023}, an intrinsic mechanism is the most likely source for the spotted regions and chromospheric activity.

\section*{Acknowledgements}

J.~Farihi is grateful to the Laboratory for Atmospheric and Space Physics at the University of Colorado Boulder, and the Kavli Institute for Theoretical Physics at the University of California, Santa Barbara, for hosting during extended visits, and to J.~S.~Pineda for an illuminating discussion on the nature of chromospheres in sun-like and low-mass stars.  The authors acknowledge the European Southern Observatory for the award of telescope time via program 105.209J.  J.~Farihi acknowledges support from STFC grant ST/R000476/1 and National Science Foundation Grant No.~NSF PHY-1748958.  S.~P.~Littlefair acknowledges the support of the STFC grant ST/V000853/1. N.~Walters has been supported by a UK STFC studentship hosted by the UCL Centre for Doctoral Training in Data Intensive Science. S.~G.~Parsons acknowledges the support of a STFC Ernest Rutherford Fellowship.  For the purpose of open access, the authors have applied a creative commons attribution (CC BY) license to any author accepted manuscript version arising.  This paper includes data collected by the {\em TESS} mission, which is funded by the NASA Explorer Program.

\section*{Data Availability}
ULTRACAM data are available on reasonable request to the instrument team, while {\em TESS} data are available through the Mikulski Archive for Space Telescopes.

\bibliographystyle{mnras}

\bibliography{../../references}
%\bibliography{references}

%

\bsp    % typesetting comment
\label{lastpage}
\end{document}